\documentclass[twocolumn,prc,showpacs,preprintnumbers,amsmath,amssymb,floatfix]{revtex4}
\usepackage{amsmath}

\usepackage{graphicx}
\usepackage{dcolumn}
\usepackage{bm}
\usepackage{ulem} 
\usepackage[usenames]{color}
\usepackage{epstopdf}
\usepackage{epsfig}
\usepackage{float}
\usepackage{subfigure}

\newcommand{\nc}{\newcommand}       
\nc{\vc}[1] {\mbox{\boldmath $#1$}} 
\nc{\del}       {\partial}              
\nc{\bra}       {\langle}               
\nc{\ket}       {\rangle}               
\nc{\bras}[1]   {\langle #1|}           
\nc{\kets}[1]   {|#1\rangle}            
\nc{\mapleft}[1]{           
 \smash{\mathop{\,          %
  \hbox to 1.5cm{\rightarrowfill}\, }\limits_{#1}}}
\nc{\beq}     {\begin{eqnarray}} \nc{\eeq}    {\end{eqnarray}}
\nc{\nn}      {\\\nonumber} \nc{\vs}      {\vspace{-0.275cm}}
\nc{\fra}    {\frac{1}{2}}
\nc{\mb}        {\mathbf}

\begin{document}


\title{Nonrelativistic nucleon effective masses in nuclear matter: BHF versus RHF}

\author{A. Li}
\email{liang@xmu.edu.cn}
\affiliation{Department of Astronomy and Institute of Theoretical Physics and Astrophysics, Xiamen University, Xiamen 361005, China}
\affiliation{State Key Laboratory of Theoretical Physics, Institute of Theoretical Physics, Chinese Academy of Sciences, Beijing 100190, China}
\author{J. N. Hu}
\email{hujinniu@nankai.edu.cn}
\affiliation{School of Physics, Nankai University, Tianjin 300071, China}
\author{X. L. Shang}
\email{shangxinle@impcas.ac.cn}
\affiliation{Institute of Modern Physics, Chinese Academy of Sciences, Lanzhou 730000, China}
\affiliation{State Key Laboratory of Theoretical Physics, Institute of Theoretical Physics, Chinese Academy of Sciences, Beijing 100190, China}
\author{W. Zuo}
\email{zuowei@impcas.ac.cn}
\affiliation{Institute of Modern Physics, Chinese Academy of Sciences, Lanzhou 730000, China}
\affiliation{State Key Laboratory of Theoretical Physics, Institute of Theoretical Physics, Chinese Academy of Sciences, Beijing 100190, China}

\date{\today}
\begin{abstract}
The density and isospin dependences of nonrelativistic nucleon effective mass ($m^*_N$) are studied, which is a measure of the nonlocality of the single particle (s.p.) potential. It can be decoupled as the so-called k-mass ($m^*_k$, i.e., the nonlocality in space) and E-mass ($m^*_E$, i.e., the nonlocality in time). Both k-mass and E-mass are determined and compared from the latest versions of the nonrelativistic Brueckner-Hartree-Fock (BHF) model and the relativistic Hartree-Fock (RHF) model. The latter are achieved based
on the corresponding Schr¡§odinger equivalent s.p. potential in a relativistic framework. We demonstrate the origins of different effective masses and discuss also their neutron-proton splitting in the asymmetric matter in different models. We find that the neutron-proton splittings of both the k-mass and the E-mass have the same asymmetry dependences at considered densities, namely $m^*_{k,n} > m^*_{k,p}$ and $m^*_{E,p} > m^*_{E,n}$. However, the resulting splittings of nucleon effective masses could have different asymmetry dependences in these two models, because they could be dominated either by the k-mass (then we have $m^*_n > m^*_p$ in the BHF model), or by the E-mass (then we have $m^*_p > m^*_n$ in the RHF model). The isospin splitting in the BHF model is more consistent with the recent analysis from the nucleon-nucleus scattering data, while the small E-mass $m^*_E$ in the RHF case as a result of the missing ladder summation finally leads to an opposite splitting behaviour.
\end{abstract}



\maketitle

\section{Introduction}

The nucleon effective mass $m^*_N$ defines the nonlocal
nature of single particle (s.p.) felt by a nucleon propagating in a
nuclear medium. It is both fundamentally important and very much related to one of the main objectives of the forthcoming new generation of radioactive beam facilities: the isospin-dependence of the nuclear force, which is crucial for understanding the properties of neutron stars, symmetry energy and the dynamics of nuclear collisions~\cite{rizzo04,farine01,li04}. However, due to the difficulties from nowadays experiments and the conflicting conclusions from different model calculations~\cite{li13}, it is hard to clarify the origins of effective masses and its density, isospin and model dependences (see \cite{lixh15} for a recent progress).

There are generally two definitions of the effective mass in the literature~\cite{jam89}. One is the so-called nonrelativistic mass $m^*/m = 1 - dV(k,\epsilon(k))/d\epsilon(k)$, which measures the nonlocality of the s.p. potential $V$ as a function of the momentum $k$ and the s.p. energy $\epsilon(k)$ from a Schroedinger-like equation. It can be decoupled as  k-mass (i.e., the nonlocality in space) and E-mass (i.e., the nonlocality in time).  The other one is the Dirac mass, which is determined by the scalar part of the nucleon self-energy and is a genuine relativistic quantity. In the present study, we focus on the first definition of the nucleon effective mass and aim to contribute a more deeper understanding of this important quantity based on the calculations and the comparisons of various most advanced nuclear many-body models. Specially, we'd like to address the isospin dependence of nucleon effective masses from the nonlocality in space and that in time, respectively, which were not discussed clearly before.

Our employed models include the nonrelativistic Brueckner-Hartree-Fock (BHF) model~\cite{book,zuo99} in combination with a microscopic three-body force (TBF)~\cite{Zuowei02prc1,Zuowei02prc2}, the relativistic Hartree-Fock (RHF)~\cite{long06plb} model with density-dependent meson-nucleon couplings, compared with the results from the Dirac-Brueckner-Hartree-Fock (DBHF) model~\cite{ma04plb,dbhf05}. For the last two models, the nonrelativistic masses are derived by rewriting the Dirac equation in a Schroedinger-type one, although they are presented in relativistic frameworks.

The BHF model can describe the equation of state (EoS) of the nuclear matter in consistent with the heavy-ion flow investigations~\cite{flow} and the observational constrains from the two recent precisely-measured heavy pulsars¡¯ masses~\cite{2mass13,2mass10}. In fact, it has been used in many studies for the structures of the neutron stars~\cite{Li11y1,Li11y2,Li11y3,Li11y4} and hybrid stars (neutron stars with quark matter in the cores)~\cite{Li15q1,Li15q2}. The detailed modelling of the BHF nuclear many-body approach is described elsewhere~\cite{book,zuo99}. Here the input bare nucleon force is the Argonne V18 two-body interaction~\cite{v18}, accompanied by a microscopic three body force constructed from the meson-exchange current approach~\cite{Zuowei02prc1,Zuowei02prc2}. It can give satisfactory nuclear matter bulk properties, which are collected in Table 1.
\begin{table}
\begin{center}
\begin{minipage}{78mm}
\caption{\small Nuclear matter bulk properties (the saturation desnity $\rho_0$, the binding energy per particle $E/A$, the symmetry energy $E_{\rm sym}$, the compression modulus $K$), obtained from the BHF+TBF model and the RHF model (PKA1) employed in the present work.}
\begin{tabular}{ccccc}
\hline
&$\rho_0 $
&$E/A$
&$E_{\rm sym}$
&$K$   \\
&[fm$^{-3}]$
&[MeV]
&[MeV]
&[MeV]  \\
\hline
BHF+TBF & 0.20  &  -14.7  & 30.6 & 226  \\
\hline
RHF     & 0.16 &  -15.8   &36.0 & 230 \\
\hline
\end{tabular}
\end{minipage}
\end{center}
\end{table}

On the other hand, the RHF model is another powerful nuclear many-body model, which was developed based on the covariant density functional theory~\cite{bouyssy87prc}. It can describe quantitatively both the finite nuclei and nuclear matter systems very well with new density-dependent meson-nucleon coupling constants proposed by Long et al.~\cite{long06plb}, and has been widely used for the studies of nuclei shell structure~\cite{long07prc,long09plb}, neutron stars~\cite{sun08prc,long12prc}, nuclei beta decay~\cite{niu13plb} and so on. The s.p. potential in the RHF model is a nonlocal quantity from the exchange term, and the corresponding k-mass and E-mass can be easily defined. There are several very successful density-dependent RHF parameter sets, such as PKO1, PKO2, PKO3~\cite{long06plb,sun08prc} and PKA1~\cite{long07prc,long08el}. All of them are fitted from the empirical properties of symmetric nuclear matter at the saturation point and the ground state properties of stable finite nuclei. The effective masses at saturation densities are $m^*_N/m_N = 0.59 , 0.60, 0.59, 0.55$ for PKO1, PKO2, PKO3, PKA1, respectively. The latest PKA1 parameter set includes an extra tensor coupling between the $\rho$ meson and the nucleon, and could describe very well the nuclear shell structure~\cite{long07prc,long08el}. Therefore in the present study we would like to use PKA1 as one representative set to discuss the effective masses for the RHF model. Its saturation properties are also collected in Table 1.

We provide the necessary formula of the effective mass and discussions of our results in Sect.~II, before drawing conclusions in Sect.~III.

----------------------------------------------------------------
\section{Formalism and Discussion}

In the nonrelativistic BHF approach, the effective mass $m^*$ is given in
\begin{eqnarray}
\frac{m^*}{m} = 1 - \frac{dV(k,\epsilon(k))}{d\epsilon(k)} = [1 + \frac{m}{k}\frac{dV(k,\epsilon(k))}{dk}]^{-1}
\end{eqnarray}
where $V(k,\epsilon(k))$ is the s.p. potential in the mean-field level with $\epsilon(k) = k^2/(2m) + V(k,\epsilon(k))$. The effective mass is actually momentum dependent, but hereafter we only consider its value at the Fermi momentum.

For the calculation of the nonrelativistic mass within the RHF and the DBHF frameworks, a Schr\"{o}dinger-type potential can be derived. For completeness we briefly introduce how it is done in the following.

The Dirac equation of a nucleon in the nuclear medium can be written as:
\begin{eqnarray}\label{dirac}
\left[\vec\gamma\cdot\vec k+m+\Sigma(k)\right]\psi=\gamma_0E\psi,
\end{eqnarray}
where $ E = \epsilon + m$ and the nucleon self-energy should be expressed by
\begin{eqnarray}
\Sigma(k)=\Sigma_S(k)+\gamma_0\Sigma_0(k)+\vec\gamma\cdot\vec k\Sigma_V(k),
\end{eqnarray}
in consistent with the rotational invariance of the infinite nuclear matter. $\Sigma_s, ~\Sigma_0, ~\Sigma_v$ are respectively the scalar, timelike and spacelike-vector components of the self-energy. In order to obtain an equivalent Schroedinger equation, the Dirac equation (Eq.~\ref{dirac}) is transformed as follows:
\begin{eqnarray}\label{dirac1}
\left[\vec\gamma\cdot\vec k+m+U_S+\gamma_0U_0\right]\psi=\gamma_0E\psi,
\end{eqnarray}
with the scalar and vector potentials defined as:
\begin{eqnarray}
U_S=\frac{\Sigma_S-m\Sigma_V}{1+\Sigma_V},~~~~~~U_0=\frac{\Sigma_0+E\Sigma_V}{1+\Sigma_V}.~~~~~~
\end{eqnarray}
This Dirac equation implies the following frequency-momentum relation,
\begin{eqnarray}
k^2+(m+U_S)^2=(E-U_0)^2,
\end{eqnarray}
which can be written in the Schrodinger-type form:
\begin{eqnarray}
\frac{k^2}{2m}+V(k,\epsilon)-\frac{\epsilon^2}{2m}=\epsilon,
\end{eqnarray}
with
\begin{eqnarray}
V(k,\epsilon)= U_S +\frac{E}{M}U_0+ \frac{U_S^2-U_{0}^2}{2m}.
\end{eqnarray}
We have omitted the $\epsilon^2/(2m)$ term in Eq.~(7) for the purpose of the present work, namely to compare the nonrelativistic effective masses from the BHF model and the RHF model. This term, generated by relativistic effects, could have evident influences on the resulting effective mass and has been thoroughly studied in Ref.~\cite{long06plb}.

The effective mass $m^*$ can be decoupled into two parts, namely k-mass $m^*_k$ and E-mass $m^*_E$ as follows~\cite{jam89,zuo05prc}:
\begin{eqnarray}
&&~~~~~~~~~~~~~~\frac{m^*}{m} = \frac{m^*_k}{m} \frac{m^*_E}{m};\\
&&\frac{m^*_k}{m} = [1 + \frac{m}{k}\frac{\partial V}{\partial k}]^{-1},~~~
\frac{m^*_E}{m} = 1 - \frac{\partial V}{\partial \epsilon},~~~~
\end{eqnarray}
which represent respectively the nonlocalities of the s.p. potential in space and that in time.

\begin{figure}
\centering
\includegraphics[width=4.cm]{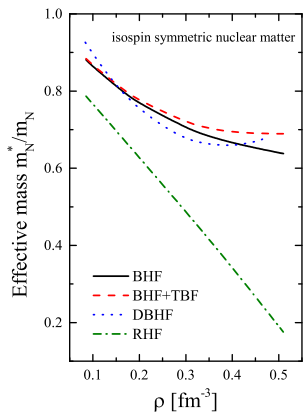}
\caption{(Color online) Nucleon effective mass $m_N^*/m_N$ as a function of the density $\rho$ for isospin symmetric matter, with the BHF model with (dashed line) or without TBF (solid line), the RHF model (dash-dotted line), to be compared with the DBHF model (dotted line).} \label{fig1}
\end{figure}

We first present in Fig.~1 the resulting nucleon effective masses $m_N^*/m_N$ as a function of the density $\rho$ for isospin symmetric matter, with the BHF model with or without TBF and the RHF model. The result from the DBHF model is also plotted here for comparison~\cite{dbhf05}. We first notice that the repulsive nature of the TBF~\cite{Zuowei02prc1,Zuowei02prc2} bring the increase of the effective mass $m_N^*$ especially at high densities. The DBHF result is similar with the BHF results, as pointed also in Ref.~\cite{dbhf05}. However, the RHF result exhibits a rapidly decreasing behaviour, which is strikingly different with the other three. This can be understood as follows: The interaction in the RHF model is mainly determined by the properties of infinite nuclear matter and finite nuclei system around the nuclear saturation density, so the constraint at high densities is missing. Furthermore, there are no high-order terms for $\sigma$ and $\omega$ mesons in the RHF model, which might suppress the contribution of scalar potential, $\Sigma_S$, on effective masses, like in the TM1 parameter set~\cite{sugahara94}. On the contrary, the ladder diagram considered in the Brueckner pair can take the strong short-range correlation into account, which would become increasingly important at high density region~\cite{dbhf05}. This short range correlation can generate a strong enhancement of the E-mass, as can be immediately seen in Fig.~2.
\begin{figure}
\centering
\includegraphics[width=4.cm]{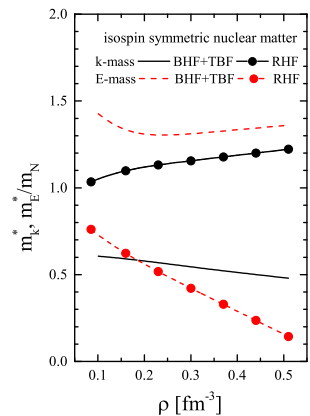}
\caption{(Color online) Density dependence of both k-mass and E-mass are compared in the BHF+TBF model and the RHF model for isospin symmetric matter.}\label{fig2}
\end{figure}

In Fig.~2, both k-mass and E-mass are compared in the BHF+TBF model and the RHF model for isospin symmetric matter. The RHF E-mass is indeed much smaller than the BHF one, and also decreases rapidly with the density. It then leads to a quick drop of the effective mass $m_N^*$ with the density (shown in Fig.~\ref{fig1}), despite the corresponding k-mass is actually larger than the BHF result. The latter is the case because the spatial nonlocality (characterized by k-mass) in the relativistic case is a combined effect from both the scalar and the vector components, $\Sigma_S$ and $\Sigma_0$, of the self-energy~\cite{jam89}. Later we will see that it is just because of the combined effect of the Brueckner ladder correlations and the mild Fock term contribution in the BHF case that results in a more consistent splitting behaviour for the nucleon effective mass $m_N^*$ with the experimentally extracted one~\cite{lixh15}.

\begin{figure}
\centering
\includegraphics[width=4.cm]{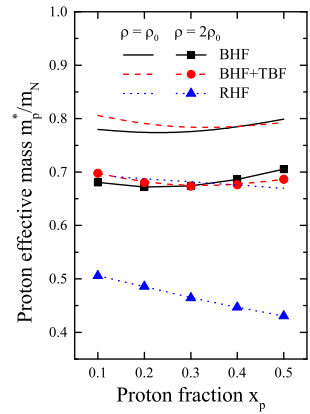}
\caption{(Color online) Proton effective mass $m^*_p/m_N$ as a function of the proton fraction $x_p$ at two densities: $\rho = \rho_0$ (shown in lines) and $\rho = 2\rho_0$ (shown in symbolled lines), for both the BHF model (with TBF in dashed lines, without TBF in solid lines) and the RHF model (dotted lines).}\label{fig3}
\end{figure}

Next, we show in Fig.~3 the results for asymmetric matter. That is, the proton effective masses $m^*_p/m_N$ as a function of the proton fraction $x_p$, for both the BHF model and the RHF model. The calculations are done for two densities: $\rho = \rho_0$ and $\rho = 2\rho_0$. We find that the results have a flatter behaviour with the TBF included in the BHF model, than the ones without the TBF. This may be seen as a suppression effect of the TBF on the change of the effective mass with the particle density. In addition, except at small $x_p$, $m^*_p/m_N$ increase with $x_p$ in the BHF model, while in the case of the RHF model, $m^*_p/m_N$ decreases monotonously with $x_p$ for the considered densities here. We can then expect that in the RHF model, with the increase of the asymmetry parameter $\beta = 1-2x_p$, $m^*_p$ increases while $m^*_n$ decreases, and we always have $m^*_p > m^*_n$, as demonstrated in the following figure.

In Fig.~4, both neutron effective mass and proton effective mass vs. the asymmetry parameter are compared in the BHF+TBF model and the RHF model for two densities: $\rho = \rho_0$ and $\rho = 2\rho_0$. As expected, the RHF model has a mass splitting feature of $m^*_p > m^*_n$ for both two cases of the densities. Only the enhanced density will result in a pronounced splitting. This is also true in the BHF case. However, in the BHF model, the splitting is opposite, namely $m^*_n > m^*_p$. The BHF result is more consistent with the recent analysis~\cite{lixh15} based on a large number of nucleon-nucleus scattering data with an isospin dependent optical model. And the dependence of the splitting on the asymmetry parameter $\beta$ is extracted as $(m^*_n-m^*_p)/m=(0.41\pm0.15)\beta$~\cite{lixh15} at normal density, to be compared with $(m^*_n-m^*_p)/m\simeq0.17\beta$ in the BHF case. We mention here that in the RHF model, the splitting at very low densities ($< 0.8\rho_0$) is actually different~\cite{long06plb} with that at high densities such as $\rho = \rho_0$ or $2\rho_0$ employed in the present work.
\begin{figure}
\centering
\includegraphics[width=4.cm]{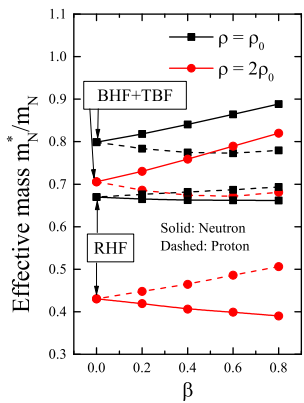}
\caption{(Color online) Asymmetry dependence of the effective mass $m^*_N/m_N$ are compared in the BHF+TBF model and the RHF model for two densities: $\rho = \rho_0$ (shown in lines with squares) and $\rho = 2\rho_0$ (shown in lines with dots). Results of neutron (proton) are displayed with solid (dashed) lines.}\label{fig4}
\end{figure}

In order to analyse further the uncertainties at different models for the isospin dependences of the neutron/proton effective masses, we show in Fig.~5 the decoupled mass splitting of both k-mass (left panel) and E-mass (right panel) in the BHF+TBF model and the RHF model for two densities: $\rho = \rho_0$ and $\rho = 2\rho_0$. From the left panel, we see that both models have the same splitting behaviour for the k-mass, namely $m^*_{k,n} > m^*_{k,p}$ for the considered two densities, although k-mass in the RHF model is somewhat larger than unity and increases with the density, while that in the BHF model is smaller than unity and decreases with the density (already seen before in Fig.~2). From the right panel, we see a much larger density effect in the RHF model for the E-mass than in the BHF model, and the RHF results are much smaller than the BHF ones. Those are consistent with previous Figs.~1-4. Also, the isospin mass splitting for the E-mass is the same in two models: $m^*_{E,p} > m^*_{E,n}$, but is opposite to the k-mass splitting: $m^*_{k,n} > m^*_{k,p}$.

The splitting of the effective mass $m^*_N$ is determined by that of the k-mass (E-mass) splitting in the BHF (RHF) model. That is to say, it is determined by the one smaller than unity. Essentially, the missing short-range correlations in the RHF model leads to a small E-mass $m^*_E$ that finally results in an opposite $m^*_N$ splitting behaviour with the experimental data and the BHF model. This may suggest that the exchange of Brueckner pairs are crucial for reproducing an experimentally derived isospin dependence for the nucleon effective mass.

\section{Conclusions}
Summarizing, we have presented a comprehensive analysis on the so-called nonrelativistic nucleon effective mass $m^*_N$ based on calculations in the latest versions of both the nonrelativistic BHF model and the RHF model. For the former one, we incorporate also the microscopic TBF, and for the latter a density dependent meson-nucleon couplings are employed. Then both of the model calculations can provide good descriptions on the experimental data of nuclear systems.
\begin{figure}
\centering
\includegraphics[width=8.cm]{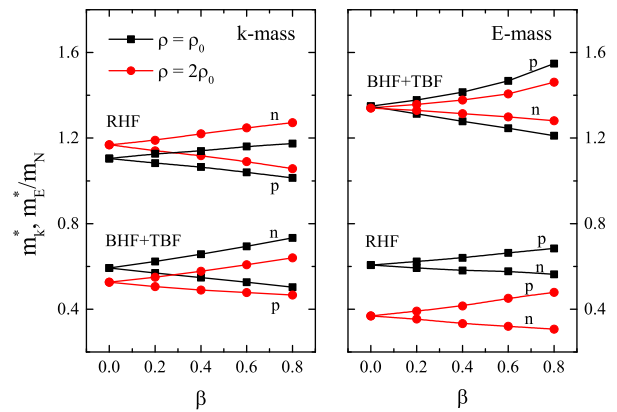}
\caption{(Color online) Mass splitting of both k-mass (left panel) and E-mass (right panel) with the asymmetry parameter are compared in the BHF+TBF model and the RHF model for two densities: $\rho = \rho_0$ (shown in lines with squares) and $\rho = 2\rho_0$ (shown in lines with dots).}\label{fig5}
\end{figure}

The nonrelativistic nucleon effective mass $m^*_N$ parameterizes the momentum dependence of the s.p. potential, and can be decoupled into two differently-defined effective masses: the usually called k-mass $m^*_k$ and E-mass $m^*_E$, which may respectively trace back to the contribution of the exchange Fock term and the Brueckner ladder correlations~\cite{dbhf05}.

We have studied in details the effects of the density and the asymmetry on the nucleon effective mass $m^*_N$, the k-mass $m^*_k$ and the E-mass $m^*_E$. We find that in the RHF model the effective mass $m^*_N$ decreases monotonously with the density, which is dominated by the rapid reduction of scalar and vector components of nucleon self-energy, while in the BHF model and the DBHF model $m^*_N$ will finally at high densities increase with the density as a result of the ladder diagram of Brueckner pair.

Furthermore, the isospin mass splittings of k-mass and E-mass have the same asymmetry dependences in both two models at considered densities. That is, $m^*_{k,n} > m^*_{k,p}$ for the k-mass $m^*_k$, but $m^*_{E,p} > m^*_{E,n}$ for the E-mass $m^*_E$ in both the BHF model and the RHF model. However, the splitting of the effective mass $m^*_N$ could be different in different models, and is determined by the one smaller than unity, namely the k-mass in the BHF case, and the E-mass in the RHF case for the considered densities in the present work. The smaller E-mass $m^*_E$ in the RHF case without the ladder summation finally leads to an opposite isospin splitting with the recent analysis from nucleon-nucleus scattering data, while the BHF model might be a more favorable model for describing the isospin dependence of $m^*_N$.

\section{Acknowledgments}
We would like to thank Prof. W. H. Long, Dr. G. F. Burgio, Dr. H.-J. Schulze for valuable discussions. The work was supported by the National Natural Science Foundation of China (Nos.11435014, 11405090, U1431107), the Major State Basic Research Developing Program of China (No. 2007CB815004), and the Knowledge Innovation Projects (Nos. KJCX2-EW-N01, KJCX3-SYW-N2) of the Chinese Academy of Sciences.

\end{document}